\documentclass[floatfix,aps,twocolumn,showpacs,nofootinbib,superscriptaddress]{revtex4}

\usepackage{graphicx}
\usepackage{epsfig}

\usepackage{amsmath}
\usepackage{amsfonts}
\usepackage{amssymb}
\usepackage{bm}
\usepackage{mathtools}

\usepackage{hyperref}
\usepackage{dsfont}
\usepackage{lipsum}

\usepackage{tikz}
\usepackage{circuitikz}
\usetikzlibrary{arrows, shapes}
\usepackage{pgfplots}
\pgfplotsset{compat=1.14}

\newcommand{\figpath}[1]{./figures/#1}
\newcommand{\colorchanges}{black}

\newcommand{\mean}[1]{\left\langle #1 \right\rangle}

\begin{document}

\author{Nahuel Freitas}
\affiliation{Complex Systems and Statistical Mechanics, Department of Physics and Materials Science,
University of Luxembourg, L-1511 Luxembourg, Luxembourg}
\author{Massimiliano Esposito}
\affiliation{Complex Systems and Statistical Mechanics, Department of Physics and Materials Science,
University of Luxembourg, L-1511 Luxembourg, Luxembourg}

\title{Characterizing autonomous Maxwell demons}
\begin{abstract}
We distinguish traditional implementations of autonomous Maxwell demons from related linear devices that were recently proposed, not relying on the notions of measurements and feedback control.
In both cases a current seems to flow against its spontaneous direction
(imposed e.g. by a thermal or electric gradient) without external energy intake.
However, in the latter case, this current inversion may only be apparent.
Even if the currents exchanged between a system and its reservoirs are inverted
(by creating additional independent currents between system and demon),
this is not enough to conclude that the original current through the system has been inverted.
We show that this distinction can be revealed locally by measuring the fluctuations of the system-reservoirs currents.
\end{abstract}

\date{\today}

\maketitle

%%%%%%%%%%%%%%%%%%%%%%%%%%%%%%%%%%%%%%%%%%%%%%%%%%%%%%%%%%%%%%%%%%%%%%%%%%%%%%%%%%%%%%%%%%%%%%
\section{Introduction}

The original notion of a Maxwell demon rests on measurements and
feedback control (that is, on the acquisition and processing of information),
to achieve local violations of the second law of thermodynamics
without violating the first law \cite{leff1990}.
The second law is restored when the system performing the measurements
and feedback is included in the analysis \cite{horowitz2014, ptaszynski2019}
(for nonautonomous setups see \cite{sagawa2008, sagawa2009, toyabe2010,  sagawa2012}).
Many model systems \cite{esposito2012, strasberg2013, strasberg2018, sanchez2019ok} as well as experiments \cite{serreli2007, bannerman2009, koski2014, koski2015, chida2017, cottet2017} have realized such traditional demons.
Recent interesting proposals have shown that similar effects
can be achieved in linear systems out of equilibrium \cite{sanchez2019, ciliberto2020}. In these proposals, a new kind of demons is characterized (`nonequilibrium demons' or `N-demons'), that cannot be interpreted in terms of measurements and feedback control, and in which the only resource is the access to a nonequilibrium distribution. Yet, a current apparently flows in a direction contrary to the one it would spontaneously flow without the action of a demon.

Our goal is to distinguish traditional formulations of Maxwell demons
from these more recent proposals.
We show that while in the former case the direction of a current is actually reversed,
in these latter cases it is in principle not possible to claim that such current inversion has taken place. What instead happens is
that the original current is supplemented by new, independent currents.
To emphasize this distinction, we take these proposals to their most
essential form by considering thermal circuits in which
on has access to the internal currents, in addition to the reservoir currents.
From this it follows that, according to the criteria accepted in \cite{sanchez2019, ciliberto2020}, a demon could be constructed using an
extremely simple and deterministic thermal circuit.
When stricter traditional criteria are used,
then those simple setups are ruled out.

We furthermore ask whether a local observer with sole access to the currents entering or
leaving the system is able to distinguish which of the two schemes is at work.
We show that this is indeed possible by studying the current fluctuations.
There is therefore an operational way to tell those two situations apart.

%%%%%%%%%%%%%%%%%%%%%%%%%%%%%%%%%%%%%%%%%%%%%%%%%%%%%%%%%%%%%%%%%%%%%%%%%%%%%%%%%%%%%%%%%%%%%%
\section{General setting}
\label{sec:setting}

\begin{figure}[ht!]
  \includegraphics[scale=.5]{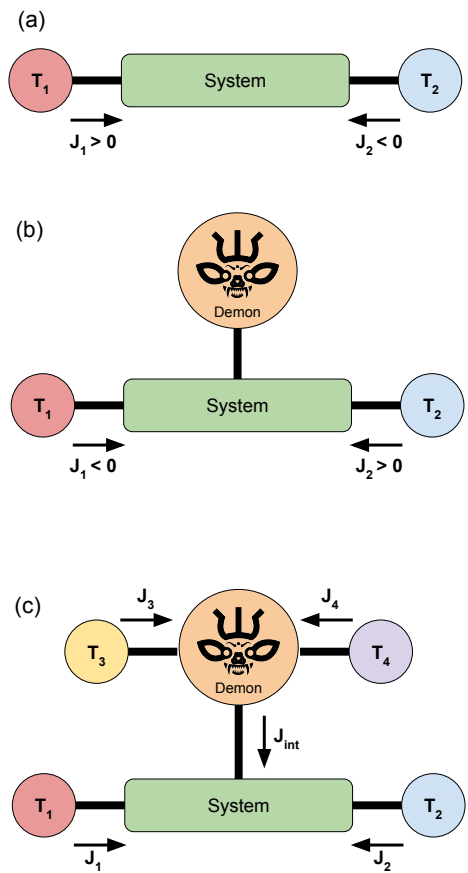}
  \caption{(a) A system interacting with only two thermal reservoirs, and thus bounded to comply with the second law. (b) When the system interacts with an additional agent (demon), then local violations of the second law might be observed. (c) However, for that to be the case, the demon must dissipate energy, which in this case is provided by a thermal gradient.}
  \label{fig:demon}
\end{figure}

We consider a system connected to two reservoirs, which can provide energy, particles, or both.
If the system is not interacting with anything else,
then it will attain a nonequilibrium stationary state with a positive entropy
production rate. For example, if the reservoirs are thermal
at temperatures $T_1$ and $T_2<T_1$ (Figure \ref{fig:demon}-(a)),
then the entropy production in the environment
is $\dot \Sigma_e = - J_1/T_1 -  J_2/T_2$,
where $J_i$ is the heat current entering the system from the $i$-th reservoir.
In stationary conditions the entropy of the system does not change so we have that
$\dot \Sigma_e$ is the full entropy production rate $\dot \Sigma$, and also,
by conservation of energy, $\mathcal{J} \equiv J_1 = - J_2$.
Then, the stationary entropy production rate can be expressed as a product
of a single current and a thermodynamic force: $\dot \Sigma_\text{st}
= \mathcal{J} \mathcal{F}$, where $\mathcal{F} = (1/T_2-1/T_1)$.
Under normal conditions, we have that the current $\mathcal{J}$ flows in the direction
of the force $\mathcal{F}$ since $\dot \Sigma_\text{st}$ is positive.
In general there might be other independent forces $\mathcal{F}_k$ and currents $\mathcal{J}_k$,
for example if the reservoirs are also able to provide or absorb particles, and thus we have
$\dot \Sigma_\text{st} = \sum_k \mathcal{J}_k \mathcal{F}_k$, but for sake of simplicity we
will focus on thermal reservoirs.

Now, if the system interacts with an external agent that is able to measure its state and manipulate
it in some way, then it might be possible to attain an stationary state with currents $J_1$ and $J_2$,
such that
the entropy production rate that one would construct based on them is actually negative (Figure \ref{fig:demon}-(b)).
If one imposes the additional condition that the external agent is a demon and is thus not allowed to exchange
energy with the system, then we still have $\mathcal{J} \equiv J_1 = - J_2$, but this time
$\dot \Sigma_\text{st}^l = \mathcal{J} \mathcal{F}$ is not necessarily positive.
The superscript $l$ in $\dot \Sigma_\text{st}^l$ indicates that this quantity is only a \emph{local entropy production rate}, compatible with the local observation of a stationary current $\mathcal{J}$, and does not take into account the entropy
production rate $\dot \Sigma_\text{st}^d$ of the demon.
A global description including the system and the demon would find
$\dot \Sigma_\text{st} = \dot \Sigma_\text{st}^l + \dot \Sigma_\text{st}^d > 0$, i.e., the negative entropy production
observed locally must be compensated by a positive entropy production in the demon.
Therefore, the demon needs to dissipate energy in order to work.
One can very generally consider that the demon obtains the energy it needs from a couple
of reservoirs imposing some gradient of temperatures or voltages.
In this way we arrive at the picture of Figure \ref{fig:demon}-(c), where
the composite system+demon is considered to interact with four independent reservoirs,
with four associated currents. Given a partition separating the degrees of freedom of
the system from those of the demon, we can also identify an internal energy current
$J_\text{int}$ between them. In this context, the following conditions
were proposed in \cite{sanchez2019, ciliberto2020} to characterize a `nonequilibrium
Maxwell demon':
\begin{itemize}
\item $\dot \Sigma_\text{st}^l = - J_1/T_1 -  J_2/T_2 < 0$
(a local violation of the second law is observed).
\item $J_\text{int} = 0$ (no \emph{net} exchange of energy between system
and demon).
\end{itemize}
We now discuss a simple example of an out of equilibrium system fulfilling these conditions.

%%%%%%%%%%%%%%%%%%%%%%%%%%%%%%%%%%%%%%%%%%%%%%%%%%%%%%%%%%%%%%%%%%%%%%%%%%%%%%%%%%%%%%%%%%%%%%
\section{An elementary example}

\begin{figure}
  \includegraphics[scale=.5]{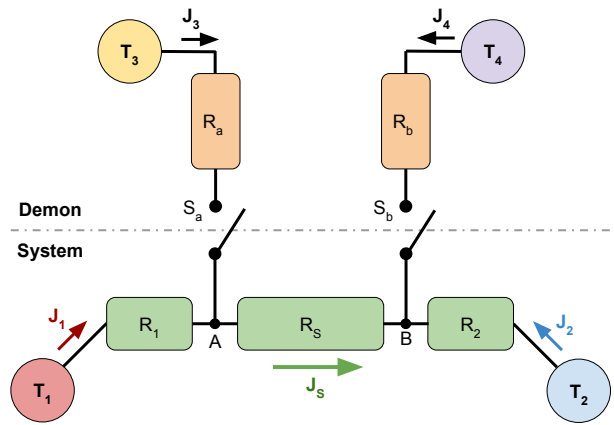}
  \caption{(a) A simple thermal circuit. The system is composed of just three linear thermal conductors of
  resistance $R_1$, $R_2$, and $R_S$, connected in series.
  The `demon' is composed of thermal conductors $R_a$ and $R_b$,
  and is connected to the system by closing the thermal switches $S_a$ and $S_b$.
  Here, black lines represent thermal conductors of negligible resistance.
  The green arrow indicates the direction of the central current when the demon is not connected (for $T_1>T_2$).}
  \label{fig:circuit}
\end{figure}

The first example in \cite{sanchez2019} involves electric and heat currents simultaneously, while the second
one only involves heat currents. Both of them are quantum systems, but this is not essential.
The example provided in \cite{ciliberto2020} is classical, and consists of a circuit with
several noisy resistors, that play the role of the thermal reservoirs.
However, as we will see, it is not necessary to consider quantum nor classical
fluctuations to achieve the conditions of the previous section (from where it
follows that correlations between currents are also not essential).
These examples can thus be further simplified.

The most elementary example that captures the essence of these proposals
is perhaps given by the thermal circuit of Figure \ref{fig:circuit}.
Here, the system is just composed of three thermal conductors
connected in series. We will assume for simplicity that these conductors are linear,
with thermal resistances $R_1$, $R_2$, and $R_S$.
This means that if a thermal gradient $\Delta T$ is applied to a conductor of resistance $R$,
the heat current through it is $J = \Delta T/R$.
The `demon' consists of two additional thermal conductors $R_a$ and $R_b$, also linear.
When the switches $S_a$ and $S_b$ are open and system and demon do not interact,
we have that the stationary reservoir currents are $J_1=-J_2=J_S>0$, since we are assuming $T_1>T_2$,
and of course the local entropy production rate $\dot\Sigma_\text{st}^l = -J_1/T_1 - J_2/T_2$
is positive. However, when the switches $S_a$ and $S_b$ are closed, stationary heat currents
$J_3$ and $J_4$ are established through the demon's conductors. In that case, by conservation
of energy at the points $A$ and $B$, in steady state conditions we must have:
\begin{equation}
\begin{split}
J_1 &= J_S - J_3 \\%= (T_1-T_2)/R_S - (T_3-T_1)/R_a\\
J_2 &= -J_S - J_4 \; .%= -(T_1-T_2)/R_S - (T_4-T_2)/R_b .
\end{split}
\label{eq:cons_energy}
\end{equation}
In turn, the net rate of energy exchange between system and demon is given
by the current
\begin{equation}
J_\text{int} = J_3 + J_4 = -J_1 - J_2 \; .
\end{equation}
The stationary currents can be computed by first noting that,
given the negligible thermal resistance of the
material connecting the central conductor $R_S$ with the conductors $R_1$ and $R_2$,
the points $A$ and $B$ can be assigned respective temperatures $T_A$ and $T_B$ (the analysis
of this kind of thermal circuits, which are common in engineering, is fully analogous to the analysis of linear electrical circuits, see for example \cite{kaviany2011}).
Thus, the different currents read $J_1 = (T_1-T_A)/R_1$, $J_2 = (T_2-T_B)/R_2$,
$J_3 = (T_3 - T_A)/R_a$, $J_4 = (T_4-T_B)/R_b$ and $J_S = (T_A-T_B)/R_S$. The stationary values
of $T_A$ and $T_B$ are then determined by the conditions in Eq. \eqref{eq:cons_energy},
which are equivalent to the following linear set of equations:
\begin{equation}
\begin{bmatrix}
\frac{1}{R_1} + \frac{1}{R_a} + \frac{1}{R_S} & \textcolor{\colorchanges}{\frac{-1}{R_S}}\\[4pt]
\textcolor{\colorchanges}{\frac{-1}{R_S}} & \frac{1}{R_2} + \frac{1}{R_b} + \frac{1}{R_S}
\end{bmatrix}
\begin{bmatrix}
T_A\\[4pt]
T_B
\end{bmatrix}
=
\begin{bmatrix}
\frac{T_1}{R_1} + \frac{T_3}{R_a} \\[4pt]
\frac{T_2}{R_2} + \frac{T_4}{R_b}
\end{bmatrix}.
\end{equation}
Solving the previous equation, we find $T_A$ and $T_B$ as a function of the resistances and the
temperatures of the reservoirs, from which all the stationary currents can be computed.

Now, the values of the temperatures $T_3$ and $T_4$ and the resistances $R_a$ and $R_b$
can be adjusted in order to fulfill the conditions of the previous
section: an apparent violation of the second law $\dot\Sigma_\text{st}^l = -J_1/T_1 - J_2/T_2 < 0$, with
no net energy exchage, $J_\text{int} =0$. However, it can be shown that under
those conditions it is not possible to invert the direction of the internal current $J_S$. To see this,
let us consider $T_A'$ and $T_B'$ to be the temperatures of the points
$A$ and $B$ when the demon is not present, and $J_1'$, $J_2'$, $J_S'$ the currents.
Since in that case we have ${J_1'=J_S'=-J_2'>0}$, it follows that
$T_A'<T_1$ and ${T_B' > T_2}$. When the demon is connected and the above conditions are achieved, we
have $J_1 = -J_2 <0$, and therefore $T_A>T_1>T_A'$ and $T_B<T_2<T_B'$. From this, we deduce that
$0 < J_S' = (T_A' - T_B')/R_S < (T_A-T_B)/R_S  = J_S$. Thus, the internal current through the system
flows in the same direction as when the demon was not connected.
Also, if a local violation of the second law is observed on the system side, then the entropy production
associated to the demon side must be positive, $\dot \Sigma^d_\text{st} = -J_3/T_3-J_4/T_4 > 0$, which together with $J_\text{int} = J_3 + J_4=0$ implies $J_4 < 0$, or equivalently,
$T_4 < T_B < T_2$. Therefore, the demon must have access to a reservoir with a temperature
lower than the minimum temperature on the system side. It can be seen that this is
a general feature of linear systems: in an arbitrary multi-terminal linear network, the reservoir
of lowest temperature is always heated at steady state (or, in other words, no linear absorption
refrigerator is possible \cite{martinez2013}). Thus, the only way to extract heat from reservoir
at temperature $T_2$ is to have access to a reservoir at a lower temperature (in this case, $T_4$). Indeed,
this is also observed in the examples in \cite{sanchez2019, ciliberto2020}.

%which in turn imply $T_4<T_2$, i.e., the demon must have access to a reservoir
%with a temperature lower than the minimum temperature of the reservoirs in contact with the system.
%This is
%also the case in the examples in \cite{sanchez2019, ciliberto2020},
%and it can be seen that it is a general feature of linear systems \cite{martinez2013}.

%\begin{figure}
%  \includegraphics[scale=.5]{\figpath{alt_circuit.pdf}}
%  \caption{In this circuit the system is composed of three thermal resistances
%  $R_1$, $R_2$ and $R_S$. The arrows in the system indicate the natural flow of currents
%  for $T_1>T_2$, when the demon is not present (again, the values are referred to the sign convention of Figure \ref{fig:demon}-(c)).
%  The temperatures of the points $A$ and $B$ are determined
%  by the stationary conditions $J_1 + J_3 - J_S =0$ and $J_2+J_4+J_S = 0$. In order to
%  invert the directions of the currents $J_1$ and $J_2$ (in such a way that $\dot\Sigma_\text{st}^l < 0$), we need to adjust $T_3$ and $T_4$ to obtain
%  $T_A>T_1$ and $T_B<T_2$. Thus, since $T_1>T_2$, we have $T_A > T_B$, and therefore
%  $J_S$ flows in the same direction as when the demon is not present. This analysis holds even if the thermal resistances are non-linear.}
%  \label{fig:alt_circuit}
%\end{figure}

Note that since the condition $J_\text{int} =0$ implies ${J_2 = -J_1}$,
a local observer that only has access to the values of $J_1$ and $J_2$ might
jump to the conclusion that these two are actually the same flow, which was reversed by the demon.
But, as this example shows, that is not always the case. Thus, it is in principle not possible to infer
what is happening inside the system just from the knowledge of the reservoir
currents and the conditions of the previous section. In sections \ref{sec:strict_criteria}
and \ref{sec:fluctuations} we discuss a more strict set of conditions to characterize
the action of a demon, and what are the corresponding signatures in the current fluctuations.

The analogy between the simple example discussed here and those provided
in \cite{sanchez2019, ciliberto2020} is clear. One difference, which is not relevant
but can obscure the comparison, is that in those examples the two new independent currents
(analogous to $J_{3/4}$ in this example) are not resolved spatially (as in Figure \ref{fig:circuit}),
but spectrally.

%%%%%%%%%%%%%%%%%%%%%%%%%%%%%%%%%%%%%%%%%%%%%%%%%%%%%%%%%%%%%%%%%%%%%%%%%%%%%%%%%%%%%%%%%%%%%%
\section{A stricter characterization}
\label{sec:strict_criteria}

The previous discussion shows that the conditions accepted in \cite{sanchez2019, ciliberto2020} to
characterize `demonic' effects actually allows for setups where
the reversion of a current is only apparent.
%, contrary to the case of a traditional demon.
However, these requirements can be
refined in order to close in on the traditional notion of a Maxwell demon,
as we will see now for systems where only heat currents are
present (like in Figure \ref{fig:demon}-(c)).
In order to do this, we first notice that all the currents involved so far are \emph{average} currents.
In particular, the condition that the demon should not provide energy to the system, $J_\text{int} = 0$,
has only been considered at the net and averaged level.
But in stricter notions of Maxwell's demons (in particular the original one \cite{leff2014}), the condition that the demon does
not provide energy to the system in question is strictly satisfied also at the fluctuating microscopic level.
This means, coming back to the setting of Figure \ref{fig:demon}-(c), that when the interaction with the
reservoirs is removed, not only the total energy of the composite system+demon is conserved, but also
separately, that of the system and of the demon.
This can of course not happen in our previous example, where system and demon continuously exchange
energy via the currents $J_3$ and $J_4$, and only the net exchange rate $J_\text{int} = J_3 +J_4$ vanish.

The condition that the internal energies of system and demon must be independent conserved quantities
(in the absence of reservoirs) has important consequences.
In particular, it implies that when reservoirs are present, there can only be two independent stationary currents \cite{rao2018}:
the one flowing through the system, that might be eventually reversed, and the one powering the demon.
Thus, under the more strict criteria, we should find that the conditions $J_1 + J_2=0$ and $J_3+J_4=0$ are always
respected in the stationary state, for any value of the intensive parameters of the reservoirs (here temperatures).
Note that in the example of the previous section there are actually three independent currents,
since the only constraint imposed by the conservation of the global energy is $J_1 + J_2 + J_3 +J_4 = 0$.
The more stringent and equivalent conditions $J_1 + J_2=0$ and $J_3+J_4=0$ can only
be achieved by specific relations between the free parameters.
Then, a global criteria to decide whether or not demonic effects in a strict sense are present is to have a setup
like the one in Figure \ref{fig:demon}-(c) where the conditions $J_1 + J_2=0$ and $J_3+J_4=0$ are always respected
(i.e., for arbitrary temperatures), but in which the currents $J_1$ and $J_2$ are anyway such that
a negative local entropy production rate $\dot\Sigma_\text{st}^l = -J_1/T_1 - J_2/T_2 < 0$ is observed.
From this, a real inversion of the current through the system can be safely concluded.
These conditions also rule out simple setups like the one of Figure \ref{fig:circuit}.
A simple mesoscopic model satisfying this strict criteria is described in the next section.

Two related comments follow.
First, we do not believe that this strict characterization is the only meaningful one.
As the one in \cite{sanchez2019, ciliberto2020} is too permissive, the one presented above is too restrictive and idealized.
Genuine demonic effects might be present also in setups where system and demon are allowed to exchange energy at the
fluctuating level. We will come back to this point in the next section and in the conclusions.
Secondly, the strict criteria involves the knowledge of the four currents $J_{1,\cdots,4}$.
A natural question to ask is whether or not it is possible for a local observer, that only has access to the values
of the currents $J_1$ and $J_2$, to confirm or discard the presence of demonic effects in a strict sense.
As we will see in Section \ref{sec:fluctuations}, such an observer might be able to infer the presence of
a spurious agent affecting the energy of the system (and thus not qualifying as a demon according to the strict criteria) by studying the current fluctuations.

%%%%%%%%%%%%%%%%%%%%%%%%%%%%%%%%%%%%%%%%%%%%%%%%%%%%%%%%%%%%%%%%%%%%%%%%%%%%%%%%%%%%%%%%%%%%%%
\section{A strict Maxwell demon model}
\label{sec:strict_demon}

\begin{figure}
  \includegraphics[scale=.6]{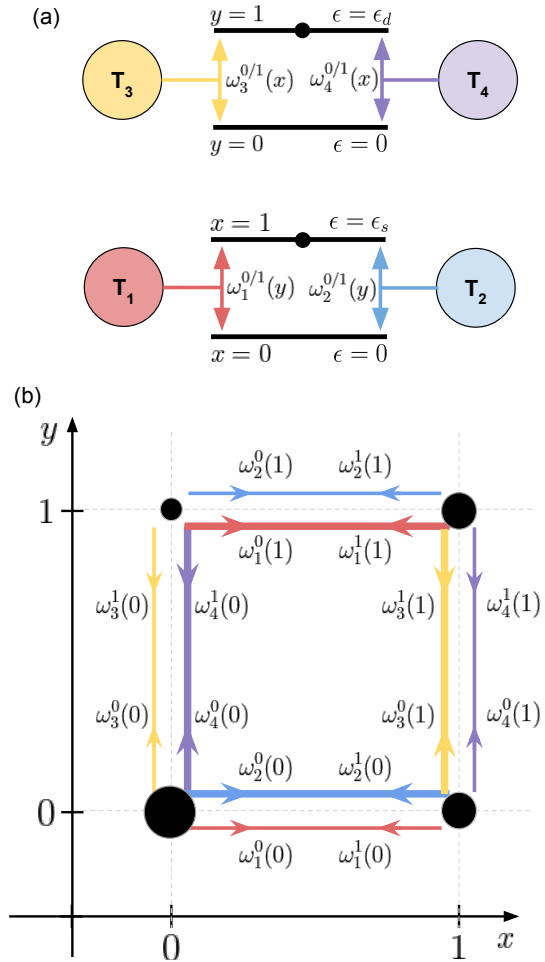}
  \caption{A mesoscopic model of an autonomous Maxwell demon satisfying the strict criteria. (a) System and demon are
  two-level systems with state dependent rates, modelling the measurement and feedback processes. (b) Global state space with possible transitions and associated rates. The size of the dots indicates the relative stationary probabilities of each state, and the thickness of the lines the relative strength of the transition rates.}
  \label{fig:strict_demon}
\end{figure}

A minimal mesoscopic model of an autonomous Maxwell demon satisfying the strict criteria is sketched in Figure \ref{fig:strict_demon}.
Both system and demon are two-level systems, with excited state energies $\epsilon_s$ and $\epsilon_d$, and state variables $x=0,1$ and $y=0,1$, respectively. Changes in the state of the system (resp. demon) are induced by interactions with thermal baths 1 and 2 (resp. 3 and 4).
Transitions $x \to \bar x$ (resp. $y\to \bar y)$ have rates $\omega_{1/2}^x(y)$ (resp. $\omega_{3/4}^y(x)$), where $\bar 0 =1$ and $\bar 1= 0$.
The dependence of the demon rates on the system state models the measurement process,
while the dependence of the system rates on the demon state models the feedback process.
By construction, measurement and feedback only happen at the kinetic level (the energies $\epsilon_{s/d}$ are state-independent constants),
and therefore in the absence of reservoirs (transitions) the energies of system and demon are independent conserved quantities.
Thermodynamic consistency is enforced via the local detailed balance conditions:
\begin{equation}
    \frac{\omega_{1/2}^0(y)}{\omega_{1/2}^{1}(y)} =
    e^{-\beta_{1/2} \epsilon_s}
    \quad
    \text{and}
    \quad
    \frac{\omega_{3/4}^0(x)}{\omega_{3/4}^{1}(x)} =
    e^{-\beta_{3/4} \epsilon_d},
\end{equation}
with $\beta_j = (k_b T_j)^{-1}$.

The measurement process works as follows. We consider
$T_3 \gg \epsilon_d/k_b \gg T_4$, and
$\omega_3^y(x=0) \ll \omega_4^y(x=0)$. Therefore, when the state of the system is $x=0$, the steady state of the demon is dominated by its interaction with the low temperature reservoir
$4$, and most of the time its state is $y=0$. When the system state is $x=1$, we consider $\omega_3^y(x=1) \gg \omega_4^y(x=1)$, so that the interaction with the hot reservoir 3 dominates and the demon state is 1 or 0 with almost equal probabilities. This establishes correlations between the system and demon states.

The feedback process works as follows. We consider $T_1>T_2$
and $\omega_1^x(y=0) \ll \omega_2^x(y=0)$ (recall that from the previous paragraph, $y=0$ indicates that $x=0$ with some degree of confidence). As a consequence, when
the system gets excited, it predominantly does so by absorbing energy from the cold reservoir 2. Finally, for $y=1$ (which implies that the system is excited, with high probability),
we consider $\omega_1^x(y=1) \gg \omega_2^x(y=1)$, and therefore
when the system decays it predominantly does so by emitting energy into the hot reservoir 1. In this way the natural heat flow between reservoirs 1 and 2 is reversed by the action of the demon. Additionally, one might consider that the measurement process is fast by taking $\omega_{3/4}^y \gg \omega_{1/2}^{x}$.

\begin{figure}
  \includegraphics[scale=.23]{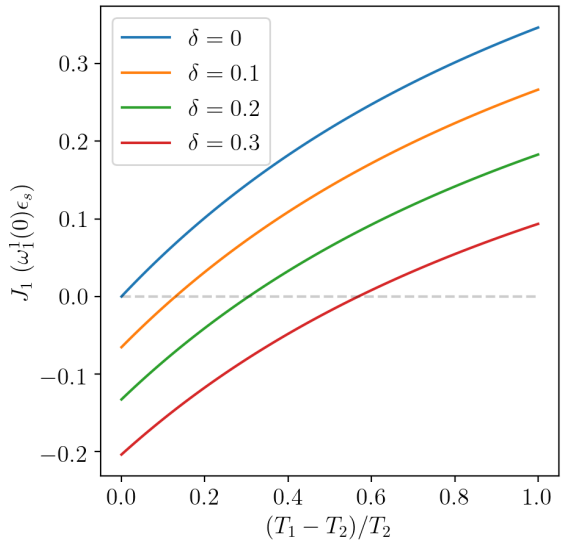}
  \caption{Current through the system as a function of the thermal bias $T_1-T_2$. The energies are $\epsilon_d = 5\epsilon_s$. The fixed temperatures are $T_2=\epsilon_s/k_b$
  and $T_{3/4}$ as given in the text. The ratios between decay rates are $\omega_2^1(0)/\omega_1^1(0) = \omega_1^1(1)/\omega_2^1(1) =
  \omega_4^1(0)/\omega_3^1(0) =
  \omega_3^1(1)/\omega_4^1(1) = 10$, and $\omega_{3/4}^1(z) / \omega_{1/2}^{1}(z) = 5$, for $z=0,1$. The other rates can be determined by the local detailed balance conditions.}
  \label{fig:results}
\end{figure}

\begin{figure}
  \includegraphics[scale=.25]{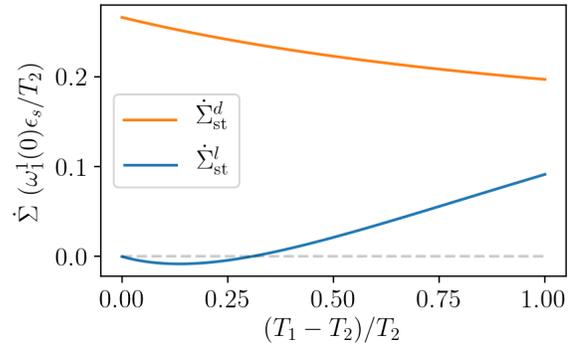}
  \caption{System and demon stationary rates of entropy production as a function of the bias $T_1-T_2$. The parameters are the same as in Figure \ref{fig:results}, and $\delta=0.2$.}
  \label{fig:entropies}
\end{figure}

Given the values of temperatures, energies, and rates, one can construct the master equation describing the stochastic
dynamics of the global system, solve for the steady state, and compute the average currents (see e.g. \cite{horowitz2014,rao2018}).
Doing that, we obtain the results of Figure \ref{fig:results}, where we show the current $J_1$
as a function of the system bias $T_1 - T_2$ (of course, we always find $J_2=-J_1$ and $J_4=-J_3$).
We show results for different values of the thermal bias powering the demon, which is parametrized
by $\delta$ according to the relations $T_3 = (1+\delta) \epsilon_d/k_b$ and $T_4 = (1-\delta) \epsilon_d/k_b$.
We see that for $\delta=0$ the current $J_1$ is always positive.
This must be the case since for $\delta=0$ the demon has no power and local violations of the second law should not be possible.
However, for $\delta>0$ we see that the current through the system can take negative values, resulting in local violations of the second law.
These negative values take place up to a maximum value of the bias $T_1-T_2$, that naturally increases as we increase the power available to the demon.
In Figure \ref{fig:entropies} we plot the stationary entropy production rate for system and demon, and we see that the entropy production in
the demon amply compensates the negative entropy production in the system, ensuring the validity of the global second law.
Finally, we note that in contrast to linear systems, in this case the demon can work even if $\min\{T_3,T_4\} \nless \min\{T_1,T_2\}$.
This is natural, since the only role of the bias $T_3-T_4$ is to power the demon, and therefore the absolute values of the temperatures
$T_{3/4}$ are not constrained by the values of $T_{1/2}$.
%In fact, the condition that the possibility of demonic effects should not be
%affected by the shift $T_{3/4} \to T_{3/4} + \Delta T$ also serves to rule out simple setups %with apparent inversion of currents.

The only purpose of this example was to provide a simple model of a demon satisfying the strict criteria. Of course, a complete model of a demon must describe the mechanism by which the rates $\omega_{1/2}^x(y)$ and $\omega_{3/4}^y(x)$ depend on the state.
As in the case in
previous and more realistic proposals \cite{strasberg2013, horowitz2014, whitney2016, sanchez2019ok}, this might involve energetic interactions between system and demon, that will not satisfy the strict criteria. However, nothing prevents those deviations from perfect energy conservation in the system to be made negligible. This is analogous to the assumption, in the original thought experiment by Maxwell, that the energetic costs associated to the measurement and feedback play no fundamental role and can in fact be neglected.

%%%%%%%%%%%%%%%%%%%%%%%%%%%%%%%%%%%%%%%%%%%%%%%%%%%%%%%%%%%%%%%%%%%%%%%%%%%%%%%%%%%%%%%%%%%%%%
\section{Detecting strict demons}
\label{sec:fluctuations}

\begin{figure}
  \includegraphics[scale=.25]{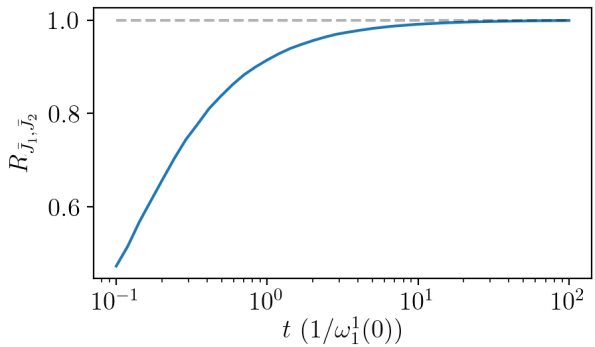}
  \caption{Pearson correlation coefficient $R_{\bar J_1, \bar J_2}$ (in absolute value) for the currents $\bar J_1$ and $\bar J_2$ as a function of the observation time $t$, for the model of Section \ref{sec:strict_demon}. The curve was obtained from stochastic trajectories generated for the same parameters of Figure \ref{fig:results} and $(T_1-T_2)/T_2=\delta=0.2$.}
  \label{fig:correlation}
\end{figure}

Let us consider a stochastic trajectory of the system internal state $\mathbf{x}_t$.
%A particular trajectory is fully characterized by its initial state $\mathbf{x}_0$, the set of transitions $\{\rho_l\}$ that took place
%up to time $t$ (each transition causes a change $\Delta \mathbf{x}_\rho$ in the system state), and the times $\{t_l\}$ at which they occurred.
The condition that the system energy must be conserved in the absence of reservoirs,
implies that when they are connected, the energy balance of the system is just:
\begin{equation}
\Delta U = Q_1 + Q_2,
\label{eq:energy_balance}
\end{equation}
where $U(\mathbf{x})$ is the internal state of the system,
${\Delta U = U(\mathbf{x}_t) - U(\mathbf{x}_0)}$ is its net change during the trajectory, and
\begin{equation}
    Q_{1/2} = \int_0^t d\tau \: j_{1/2}(\tau)
\end{equation}
are the total amounts of heat interchanged with the reservoirs.
In the last equation, the currents $j_{1/2}(\tau)$ are functionals of the trajectory.
%that vanish everywhere except for $\tau = t_l$, the times of the jumps.
At steady state, the heats $Q_{1/2}$ are time extensive, while the energy difference
$\Delta U$ is bounded. Thus, from Eq. \eqref{eq:energy_balance}, in the limit of long times we have
\begin{equation}
    \lim_{t\to\infty} (\bar J_1 + \bar J_2) =  \lim_{t\to\infty} \Delta U/t = 0,
\end{equation}
where $\bar J_{1/2} = t^{-1} Q_{1/2}$ are the mean currents during the trajectory.
This means that for long times the system currents become perfectly anticorrelated.
Actually, one must be more careful, since in that same limit the mean currents become deterministic variables, with negligible fluctuations.
However, while $\bar J_1 + \bar J_2$ scales as $t^{-1}$, the standard deviation $\sigma_{k} = \sqrt{\mean{(\bar J_k - \mean{\bar J_k})^2}}$
of each current scales as $t^{-1/2}$ (this scaling
can be understood as a consequence of the central limit theorem).
Therefore, there is a long time regime for which the fluctuations of $\bar J_1$ and $\bar J_2$
are non vanishing and almost perfectly anticorrelated. This is illustrated in
Figure \ref{fig:correlation} for the model of the previous section, where we show the Pearson correlation coefficient $R_{\bar J_1, \bar J_2} = \mean{(\bar J_1 - \mean{\bar J_1})(\bar J_2 - \mean{\bar J_2})}/(\sigma_{1} \sigma_{2})$ as a function of the observation time $t$.
This signature is a consequence of the fact that the energy of the system is a conserved quantity, independent of the energy of the demon.
If that is not the case, then the energy balance for the system is:
\begin{equation}
\Delta U = Q_1 + Q_2 + Q_\text{int},
\end{equation}
where $Q_\text{int}$ is the energy provided to the system by the demon during a trajectory.
In principle, this quantity has fluctuations that are independent of the fluctuations in $Q_{1/2}$.
Furthermore, even if the average of $\bar J_\text{int} = t^{-1} Q_\text{int}$ vanishes
(which is the condition in \cite{sanchez2019, ciliberto2020}),
its fluctuations also scale as $t^{-1/2}$, and therefore are comparable
to the fluctuations of $\bar J_{1/2}$ (this scaling is also a consequence of the central limit theorem). Thus, a local observer that is able to measure
the currents $\bar J_{1/2}$ and fails to observe a perfect anticorrelation between them (for long times),
can conclude that the system in question is exchanging energy with an additional agent,
apart from the reservoirs 1 and 2. In this way, demons satisfying the strict criteria can be locally
distinguished from those setups in which the inversion of the current through the system is only apparent.

%%%%%%%%%%%%%%%%%%%%%%%%%%%%%%%%%%%%%%%%%%%%%%%%%%%%%%%%%%%%%%%%%%%%%%%%%%%%%%%%%%%%%%%%%%%%%%
\section{Conclusions}

We discussed qualitative differences between the traditional notion of autonomous Maxwell demons and more recent proposals displaying similar effects \cite{sanchez2019, ciliberto2020}.
We showed that the characterization of `demonic' effects put forward in those proposals allows for extremely simple setups
where an interpretation of the observed effects in terms of currents flowing against a gradient (a hallmark of Maxwell demons) does not hold. We proposed a stricter set of conditions
aiming at discarding those simple setups, while still capturing the original notion of a Maxwell demon.
We also showed that our criteria can be tested operationally by studying currents fluctuations.
To highlight the essence of our argument, we focused on systems that are only in contact with heat reservoirs.
A more complete and realistic characterization of autonomous Maxwell demons should consider systems with coupled currents of different nature such as thermoelectric systems \cite{strasberg2013, horowitz2014, koski2014, koski2015, whitney2016, chida2017, cottet2017, sanchez2019ok}.
In these cases, the demon task is to invert an electric current.
To do so, energy (but not charge) flows between the system and the demon may be allowed.
%to fluctuate.
However, perfect correlations in the long time measurements of input and output electrical currents should be ensured. \textcolor{\colorchanges}{
This is indeed the case in all the proposals and experimental realizations
cited above, as can be seen by just noticing that the charge in the demon side is conserved}.

In general, our strict criteria might be relaxed by allowing departures from perfect correlations, provided that these
departures (which indicate that the system energy or charge is not truly conserved) are not large enough to explain
the observed local violation of the second law. But such arguments would need to be quantified.

%%%%%%%%%%%%%%%%%%%%%%%%%%%%%%%%%%%%%%%%%%%%%%%%%%%%%%%%%%%%%%%%%%%%%%%%%%%%%%%%%%%%%%%%%%%%%%
\section{Acknowledgements}

We thank Sergio Ciliberto, Robert S. Whitney, Janine Splettstoesser, and Rafael Sánchez for useful comments on the manuscript.
We acknowledge funding from the European Research Council, project NanoThermo (ERC-2015-CoG Agreement No.681456),
and from the FQXi foundation, project ``Information as a fuel in colloids and superconducting quantum circuits'' (FQXi-IAF19-05).

%%%%%%%%%%%%%%%%%%%%%%%%%%%%%%%%%%%%%%%%%%%%%%%%%%%%%%%%%%%%%%%%%%%%%%%%%%%%%%%%%%%%%%%%%%%%%%

\bibliographystyle{unsrt}
\bibliography{references.bib}

\end{document}